\documentstyle[preprint,aps]{revtex}
\widetext
\draft
\tighten

\begin{document}

\title{Coulomb interaction does not spread instantaneously}

\bigskip

\author{{\bf  Rumen I. Tzontchev, Andrew E. Chubykalo and Juan M.
Rivera-Ju\'arez}}

\address {Escuela de F\'{\i}sica, Universidad Aut\'onoma de Zacatecas \\
Apartado Postal C-580\, Zacatecas 98068, ZAC., M\'exico\\
e-mails: {\rm rumen@ahobon.reduaz.mx} and {\rm andrew@ahobon.reduaz.mx}}

\date{\today}

\maketitle

%\date{December 12, 1995}

\baselineskip 7mm
\bigskip
\bigskip
\bigskip
\bigskip

%\pacs{PACS numbers: 03.50.-z, 03.50.De}

% \clearpage
\begin{abstract}
The experiment is described which shows that Coulomb interaction spreads
with a limit velocity and thus this kind of interaction cannot be
considered as so called ``instantaneous action at a distance"
\end{abstract}

\section{Introduction}

As shown in previous works by one of the authors of this article [1-4],
instantaneous action at a distance is the direct consequence of generally
accepted classical electrodynamics. Particularly in the aforementioned
works it was shown that the Coulomb field (unlike a so called {\it free}
field) is spread instantaneously. On the other hand theories exist which
affirm that any kind of electric field spreads with a limited rate, but
these theories require significant modification within classical
electrodynamics [5]. However, one cannot choose theoretically one theory
over another; it must be done through experiment. Thus, the propose of
this present work is to verify experimentally if the Coulomb field is
really spread instantaneously or not.

In the following section we describe an experiment which provided us with
appropriate framework for verifying the instantaneous spread.

\section{Experimental equipment}

In order  to determine the velocity of the Coulomb interaction, a Coulomb
electric  field generator, which rapidly changes its magnitude, and
antennas which can register this electric field, are necessary.

\subsection{The generator}

The  generator consists of two hollow metallic spheres $a$ and $b$,
connected electrically by means of a plasma discharger  and discharge
cable (fig.1):
$$$$
Fig1. {\small Coulomb electric field generator (regular configuration)}
$$$$

The  radius of each sphere is $R$, the distance
between the centres of the spheres is equal to $l$. Let us investigate the
field at point $A$, which is on a line which passes through the centres of
both spheres and which is at a distance $L$ from the centre of sphere $b$.
Let us call this line the ``experimental axis".

Sphere $b$ is charged  electrically with a positive charge as far as the
break-down voltage of the discharger. Then, a spark flies through the
discharger, a spark which is a very brief plasma cord of little electrical
resistance.  As we can see, both spheres are electrically connected during
the spark by means of the discharge cable alone. Thus begins a process of
discharge, during which a part of the electric charge initially
accumulated on sphere $b$ oscillates between sphere $a$ and sphere $b$
with a high frequency. The distance between this mobile part of the charge
and point $A$ changes with the same frequency. Consequently, the electric
field at point $A$ also oscillates according to the Coulomb law. This field
has colinear direction at the experimental axis.

It is important to note
that because of the cilindrical symmetry of the experimental construction
at point $A$, the electric field is not perceived due to an
electromagenetic transversal wave, regardless of the values of the
distances $l$ and $L$.

In our examinations further on we will use the
quasi-stationary approximation.  This approximation can be correctly
applied only if the length of the discharge cable,  the radius of the
spheres and all distances in our experiment are less than the wavelength
of the proposed field.  All our calculations will be carried out supposing
that ``instantaneous action" exists, that is, the velocity of propagation
of the Coulomb electric field is infinite. In this case the wavelength
will have an infinite value.  In case it turns out that the electric field
is propagated with a finite velocity and if the use of the
quasi-stationary approximation is not correct, our calculations can only
be considered as the first aproximation in the description of the process.

From an electric point of view, both spheres
$a$ and $b$ are condensers. Their capacitance is the sum of each  sphere's
own capacitance and the capacitance in common with the other sphere:
$$
C= C_{own} + C_{a,b}
$$
The  discharge cable can be presented as an induction $L_d$ and a
resistance $R_d$  connected in series. When the discharger is in the state
of conduction it can  be substituted by a short circuit. Consequently, the
variable field  generator can be presented as a dipole (fig.2):

$$$$
Fig.2. {\small Electric diagram equivalent to the electric field
generator.}
$$$$
$q_a(t)$  and $q_b(t)$ are the electric charges
corresponding to spheres $a$ and $b$ and $U_a(t)$ and $U_b(t)$ are the
corresponding potentials.  Applying the second law of Kirchoff for this
circuit and taking into consideration that each sphere is in itself an
equipotential surface, we obtain:
\begin{equation} R_d
i(t)+L_d\frac{di(t)}{dt}-\frac{1}{C}\left[q_a(t)-q_b(t)\right]=0,
\end{equation}
where $i(t)$  is the current which passes through the discharge cable.

The  process begins by electrically charging sphere $b$ as far as the
break-down  voltage of the discharger $U_{b0}$. Then, when the spark
appears on the surface  of sphere $b$ a charge of
$$
Q_0 = CU_{b0}
$$
has accumulated.

Just as during  the process of discharge there is no charge leak, and the
capacitance of the discharge cable is not appreciable, it is always
\begin{equation}
q_a(t) + q_b(t) = Q_0.
\end{equation}
From equations (1)   and (2), for the charge $q_b(t)$ we obtain the
following differential equation:
\begin{equation}
\frac{d^2q_b(t)}{dt}+\frac{R_d}{L_d}\,\frac{dq_b(t)}{dt}+
\frac{2}{L_dC}q_b(t)=\frac{Q_0}{L_dC}
\end{equation}
with the initial conditions in the instant $t = 0$:
$$
\left.\frac{dq_b(t)}{dt}\right|_{t=0}=0; \qquad \left.q_b(t)\right|_{t=0}
=Q_0.
$$
The solution to this differential equation is:
\begin{equation}
q_b(t)=q_me^{-\beta t}\cos(\omega t+\alpha)+\frac{Q_0}{2},
\end{equation}
where
$$
\beta=\frac{R_d}{2L_d}; \;\; \omega=\sqrt{\omega_0^2-\beta^2};\;\;
\omega_0=\sqrt{\frac{2}{L_dC}};\;\; \tan\alpha=-\frac{\beta}{\omega};\;\;
q_m=\frac{Q_0}{2\cos\alpha}.
$$

For the electric charge of the sphere $a$ we obtain:
\begin{equation}
q_a(t)=Q_0-q_b(t)=\frac{Q_0}{2}-q_me^{-\beta t}\cos(\omega t+\alpha).
\end{equation}
The electric current discharge $i(t)$ is given by the expression:
\begin{equation}
i(t)=-q_m\omega_0e^{-\beta t}\sin\omega t.
\end{equation}
The potential of  the electric field at point $A$ will be the sum of the
potentials generated by the charges $q_a(t)$ and $q_b(t)$
\begin{eqnarray}
\varphi_e(L,t) & = & \frac{q_a(t)}{4\pi\varepsilon_0(L+l)}+
\frac{q_b(t)}{4\pi\varepsilon_0L} = \nonumber\\
& & =
\frac{Q_0}{8\pi\varepsilon_0}\,\frac{2L+l}{L(L+l)}+
\frac{q_m}{4\pi\varepsilon_0}\,\frac{l}{L(L+l)}\,e^{-\beta t}
\cos(\omega t+\alpha).
\end{eqnarray}
As we pointed out  earlier, in the case of the existence of a finite
velocity of the  propagation of the Coulomb interaction, the application
of the quasi-stationary  approximation can be  not correct and the
processes in the generator may only be  described qualitatively. However,
the basic conclusion of the consideration  made is correct: there are
rapid spatial oscillations of the electric charge along the axis of the
experiment that produces a rapid change of electric field  at the point
$A$.

\subsection{The antenna}

The  antenna is a metallic hollow sphere of radius $r$, connected by means
of a cable to the earth (fig.3).
$$$$
Fig.3. {\small Electric antenna.}
$$$$

It  is accepted that the earth has a potential equal to zero. Thus, the
potential on the surface of the antenna will always be equal to 0 due to
the connection to the earth.  When there is no external electrical field
the sphere of the antenna is not electrically charged.  Let us suppose
that there is an external electrical field whose source is sufficiently
far away from the sphere so that we can accept that the potential of the
external field is not altered considerably in the place of the sphere.
Then we can talk of the potential in the region on the sphere of the
antenna $\varphi_{e,a}(t)$. In this case, there is charge $q(t)$ of such
magnitude on the surface of the sphere that equation (8) is fulfilled:
\begin{equation}
\varphi_{e,a}(t)+\varphi_{i,a}(t)=\varphi_{e,a}(t) +\frac{q(t)}
{4\pi\varepsilon_0r}=0,
\end{equation}
where $\varphi_{i,a}(t)$ is  the potential on the surface of the antenna
originated by charge $q(t)$, which comes from the earth. Therefore:
\begin{equation}
q(t)=-\varphi_{e,a}(t)\,4\pi\varepsilon_0r.
\end{equation}
When the electric field changes, charge $q(t)$ changes. This means
that the electric charge passes through the connecting cable between the
antenna and the earth, in other words, electric current passes through the
connecting cable:
\begin{equation}
i_a(t)=\frac{dq(t)}{dt}=-4\pi\varepsilon_0r\,\frac{d\varphi_{e,a}(t)}{dt}.
\end{equation}
The current $i_a(t)$ can be recorded.

\subsection{Experimental configuration}

The  Coulomb electric field generator and two antennas are at a distance
$F$ from each other, as shown in fig. 4:
$$$$ Fig.4. {\small The main
diagram of the experimental configuration.}
$$$$
The  distance from the
centre of antenna 1 to the centre of sphere $b$ of the generator is
$L$. In the cables, which connect the antennas to earth, small $R_{50}$
magnitude resistors are included, which do not affect the working of the
antennas. The voltage fall on these resistors is proportional to the
current and is supplied at both input points to a rapid digital
oscilloscope. In this way, the currents passing through the cables are
recorded.

The discharge
in the discharger gives rise to an oscillatory displacement of
electric charge along the experimental axis. For this reason, a
change in the Coulomb electric field will be emitted from the
generator through the same axis. This change, reaching antennas 1
and 2 will cause the current to flow through the $R_{50}$ resistors and
thus, a $U_R$ voltage fall upon them.  In the case of  ``instananeous
action", when the Coulomb electric field appears simultaneously in
the whole space, this $U_{R1}(t)$ voltage fall for antenna 1 satisfies the
following differential equation:
\begin{equation}
\frac{dU_{R,1}(t)}{dt}+\frac{1}{4\pi\varepsilon_0rR_{50}}\,U_{R,1}(t)
=\frac{d\varphi_{e,1}(t)}{dt},
\end{equation}
where $\frac{d\varphi_{e,1}(t)}{dt}$      is  the first derivate of the
potential of the Coulomb electric field in the region of antenna 1,
originated by the generator.  The inicial condition $U_{R,1}(0)=0$ gives
us the following solution of this equation:
\begin{equation}
U_{R,1}(t)=\frac{D}{(\beta-B)^2+\omega^2}\left[-\omega e^{-Bt}
+\omega e^{-\beta t}\cos\omega t+(\beta -B)e^{-\beta t}\sin\omega t
\right],
\end{equation}
where
$$
D=\frac{q_m\omega_0}{4\pi\varepsilon_0}\,\frac{l}{L(L+l)};\qquad
B=\frac{1}{4\pi\varepsilon_0rR_{50}}.
$$

The voltage  fall $U_{R,2}(t)$  corresponding to antenna 2 obeys the same
law and the difference  is only in the value of the coefficient $D$. For
antenna 2:
$$
D=\frac{q_m\omega_0}{4\pi\varepsilon_0}\;\frac{l}{(L+F)(L+F+l)}.
$$

The tension $U_{R,1}(t)$ has been supplied to channel 1 of the
oscilloscope by means of a coaxial cable, and the tension $U_{R,2}(t)$ -
at channel 2.  Consequently,  {\bf if ``instantaneous action" exists, two
impulses of equal form and different amplitude should be visible on the
display of the oscilloscope, which appear simultaneously and develop in a
parallel fashion in time}. We assume that the air and coaxial cable are
linear structures which possess no noticeable dispersion for the
frequencies used.  The other possibility is that the change in the Coulomb
electric field is propagated at a finite velocity, that is, a wave exists.
So, the front of the wave produced by the Coulomb electric field will
arrive at antenna 1 first and then at antenna 2 after a certain interval
of time $\Delta t$.  Consequently, two impulses displaced in time $\Delta
t$ will be seen on the display of the oscilloscope, which will obey the
equation:
$$ \Delta t=\frac{F}{v},
$$
where $v$ is the  velocity of
propagation of the wave. The interval can be measured experimentally and
the velocity of the front of the wave $v$ can be obtained, which by its
essence, is group velocity.

\section{Practical experimental equipment}

The  Coulomb electric field generator consists of two standard Van de
Graaf generators with the radius of the balls equal to 10 cm. One of the
generators becomes electrically charged during the experiment, and we
shall call it ``active". The other, which is only used for receiving a part
of the charge of the active generator through the discharger and the
discharge cable, we shall call ``passive". The generators are elevated on
insulating supports until the centre of the balls reaches a height of 1.7m
with reference to the earth's surface. The distance between the centres of
both balls is 3 m.  The antennas are spheres of a radius of 9.5 cm, whose
centres can be found 1.7 m from the earth's surface. The centre of antenna
1 is to  be found at a fixed distance of 0.5 m from the centre of the ball
of the Van de  Graaf active generator, and the distance between the
centres of both  antennas varies. Fig. 5 gives an overview of the
practical experimental equipment:

$$$$
Fig.5. {\small Practical experimental equipment (overview).}
$$$$

Antenna  1 is slightly displaced from the experimental axis so as not to
obstruct the direct visibility between the Van de Graaf and antenna 2
generators. If this does not happen, antenna 1 will behave like a screen
and the signal from antenna 2 decreases drastically.  Both antennas are
connected at the input of the oscilloscope by means of two high frequency
coaxial cables with the characteristic resistance of 50  Ohms. Antenna 1
is connected to channel 1 and antenna 2 to channel 2.  Each cable is
impedance balanced on both sides. The lengths of the cables are equal,
with an uncertainty of 5 mm.  The noise at the input of the oscilloscope
has a value of 10 mV$_{p-p}$.  The measurement sensitivity of the temporal
intervals is 0.3 ns.  The signals which are due to the effects of
the apparatus (signal penetration from one of the channels of the
oscilloscope to the other, bad earth contact, signal penetration through
the power supply cables, etc.) and inductions in the coaxial cables have
already been measured. With this purpose, an experiment has been carried
out, in which antenna 1 has approached to a maximum the active Van de
Graaf generator maintaining constant the other variables of the
experiment; the sphere of antenna 2 has been disconnected from the coaxial
cable. The signal in channel 2 of the oscilloscope in the case of various
distances bewteen the end of the coaxial cable corresponding to antenna 2
and the active Van de Graaf generator has been measured. The result is
that in all cases, the signal obtained has a value below $0.5\%$ of the
useful signal (with the sphere of antenna 2 connected). An analogous
experiment has been done with antenna 1, commuting the exploration of the
oscilloscope to antenna 2. The result was similar. This allows us to
exclude corrections in order to avoid apparatus effects.

The temporal
symmetry of the antennas of the experimental equipment has been verified.
The defect of the aforementioned symmetry can be obtained from a different
length of the coaxial cables or an assymetry of both channels of the
oscilloscope. With this end, both antennas are disposed symmetrically
side-by-side at a distance equal to that of the active Van de Graaf
generator. The delay of the front flank of the impulse of channel 1 with
reference to the front flank of the impulse of channel 2 in the case of
various combinations of the sensitivity of the vertical amplifiers of the
oscilloscope has been determined. The result is that the delay in all
cases is below 0.3 ns. For this reason we have considered that, in the
following measurements and calculations, the assymetry shown in time is
equal to 0.3 ns.

The
experiments were carried out in the following manner:  Antenna 2 is placed
at a certain distance $F$ from antenna 1. The oscilloscope is put to work
in the framework of the exploration of single firing with a
sinchronization along the negative front flank of the impulse of channel
1. One waits until the moment of the spontaneous jump of the spark between
the electrodes of the discharger. The information of the impulses of
antennas 1 and 2 is recorded in the memory of the oscilloscope. Then the
analysis of the information begins.

It has been predicted theoretically
and demonstrated experimentally that the first flanks of impulses from
antennas  1 and 2 are negative and have the same form. For this reason, by
means of  an adjustment of the sensitivity of the vertical amplifiers of
both channels  of the oscilloscope, its parallelism is reached (fig.6) and
then the delay $\Delta t$     of the impulse flank of antenna 2 with
reference to the impulse flank of antenna 1 is measured.

$$$$
Fig.6. {\small Disposition of the antennas'  impulses on the display of the
oscilloscope.}
$$$$

This investigation  has been repeated for different distances between
antennas 1 and 2 in the interval of 0.50 m to 1.50 m spaced
uniformly 0.10 m. For each value of distance the measurement has
been reiterated 20 times.  With the purpose of proving what
influence the power supply cables and the oscillations (which are
gradually propagated along them) have, a
complete cycle of investigations has been
carried out, with the configuration shown in
fig.7:

$$$$
Fig.7. {\small Coulomb electric field generator  (configuration with a turn of
$180^{\circ}$).}
$$$$

The Coulomb electric field  generator is turned $180^{\circ}$ on the
horizontal plane, and with this the passive and active Van de Graaf
generators have changed places. At the same time, the disposition
of the power supply cables and the block of antennas is
unalterable. According to the theoretical description, the front
flanks of the impulses of channels 1 and 2 in this case should
invert their sign from negative to positive and the impulses of
both channels should maintain sameness of form. This is observed
in practice and the results of the investigation for $\Delta t$  do
not differ significantly from the results obtained using the basic
configuration.  A basic parasite factor is the transversal
electromagnetic wave which is emitted from the Coulomb interaction
generator and is reflected on the earth's surface. It is slightly
different from the Coulomb interaction by the sign of its first
front and by the sign of the flank of the first impulse which can
be seen on the oscilloscope's display respectively. In the case of
the basic configuration (fig.5), the first front of the Coulomb
interaction is always negative, while the first front of the
transversal electromagnetic wave is positive. This can be
explained using the Electrodynamic theory in the case of the
construction shown of the Coulomb electric field generator [6].  In
order to separate the Coulomb interaction from the signal
originated by the transversal electromagnetic wave, the delay
observed from the first front of the transversal electromagnetic
wave with respect to the first front of the Coulomb interaction is
of assistance. This additional time is that needed by the
transversal electromagnetic wave in order to arrive at the earth's
surface and then to the antenna, while the Coulomb interaction is
propagated along a straight line.  With the purpose of decreasing
additively the amplitude of the transversal electromagnetic wave,
the discharge cable was electrically and magnetically screened.
The  aformentioned experimental facts oblige us, in order to avoid the
influence of the transversal electromagnetic wave on the experimental
results, to work with a maximum distance of 1.5 m between the antennas.
In the  absence of "instantaneous action", it is necessary to determine
which part of our measuring signal is due to the transversal
electromagnetic wave and which part is due to the lonitudinal component of
the electric field. Our antennas react to the potential of the electric
field and for this reason cannot distinguish both components from the
electric field. With the purpose of separating them, screens have been
used which considerably decrease the intensity of the transversal
electromagnetic wave without altering to a relevant degree the amplitude
of the longitudinal component. Two types of screens have been used: a
metallic mesh and a thin layer of aluminium.  In order to reject the
influence of the waves reflected in the varying objects, the antenna was
placed in a thick aluminium cylinder with one end sealed and the other
open, connected to the earth. The opening of the cylinder was directed
towards the Coulomb interaction generator. The screens were placed
covering this opening.  The metallic mesh is made of iron wire with a
diameter of 1mm and the mesh measures 5 mm $\times$ 5 mm. Its
effectiveness as armour for the transversal electromagnetic wave with the
frequency obtained by us for 9.2 MHz is greater than 50 dB[7]. This means
that the signal due to this wave decreases considerably and now barely
affects the results of the experiment. In this case, we register a
decrease of our signal of just 0.5 dB. Because of this, we can conclude
the practically all of our signal comes from the logitudinal component of
the electric field generated by the Coulomb interaction generator.  We
repeated the same experiment using a 0.02 mm thick aluminium sheet as a
screen. Its effectiveness  as armour for the transversal electromagnetic
wave of the aformentioned  frequency is greater than 80 dB[7]. The
decrease in the signal is just 1 dB,  and this result confirms the
previous conclusion that our signal from the  antenna is only due to the
longitudinal component of the electric field.

\section{Experimental results}

The results of the experiment are presented in fig.8.
$$$$
Fig.8. {\small Relation between the delay $\Delta t$ between  the impulses
of both antennas and the distance $F$ between the antennas.}
$$$$

The distance $F$ between  antenna 1 and antenna 2 has been traced along the
horizontal axis. The delay  $\Delta t$ of the signal of antenna 2 with
reference to antenna 1 has been traced on  the vertical axis. The
uncertainties of the measurement results have also  been marked on the
distance axis and the time axis.

The straight line, which is  presented in the drawing, expresses the
empirical linear relation,  which relates distance $F$ to the delay
$\Delta t$.  This has been determined by means  of the minimum square
method.  The line, which corresponds the the speed  of light $c =
3.00\times 10^8$ m/s cannot be differentiated from the line  shown by
the limited exactness of the representation on the drawing.

Following a statistical treatment  one obtains, that with a reliable
probability $P = 0.95$ the group velocity of the propagation of the
Coulomb interaction is found in the interval [8]:
$$
v = (3.03\pm  0.07)\times 10^8\; {\rm m/s}
$$

We would like to call this type of field propagation ``{\it Coulomb
waves}", which, of course, has no analogy with the usual electromagnetic
waves.

\section{conclusions}

P.A.M. Dirac wrote ([9], p. 32): ``{\it As long as we are dealing only
with transverse waves, we cannot bring in the Coulomb interactions between
particles. To bring them in, we have to introduce longitudinal
electromagnetic waves and include them in the potentials $A_\mu$.}"
It is known, however, that generally accepted Maxwellian electrodynamics
forbids the spreading of {\it any} longitudinal electro-magnetic
perturbation in vacuum. So Dirac writes ([9], p. 32): ``{\it The
longitudinal waves can be eliminated by means of a mathematical
transformation...}". But after this transformation one gets the theory
in which a charge is  {\it always} accompanied by the Coulomb field around
it, i.e. ([9], p. 32) ``{\it Whenever an electron is emitted, the Coulomb
field around it is} {\tt simultaneously} {\it emitted, forming a kind of}
{\tt dressing} {\it for the electron. Similarly, when an electron is
absorbed, the Coulomb field around it is} {\tt simultaneously} {\it
absorbed}".

In accord with the above it was shown [1-4] that the Coulomb interaction
``works" {\it instantaneously}, if one accepts that basic equations of
classical electrodynamics are right. On the other hand our experiment
shows that the Coulomb interaction, at least, does not spread
instantaneously.  The proper inference from this experiment is that the
Coulomb interaction cannot be considered as so called ``instantaneous
action at a distance" and, in turn, the basic equations of classical
electrodynamics {\it can be incomplete} and moreover, their application is
even limited in classical electrodynamics.

\end{document}